\documentclass[a4paper,12pt]{article}

\usepackage{jheppub} 

\usepackage[T1]{fontenc} 
\usepackage{bm,latexsym,amsmath,amssymb,amsfonts,mathtools,mathrsfs}
\usepackage{comment}
\usepackage{ulem}
\usepackage{color}

\usepackage{bm}
\usepackage{hyperref}
\usepackage{framed}
\usepackage{subfigure}
\usepackage{hyperref}   
\usepackage{graphicx}

\expandafter\let\csname equation*\endcsname\relax 
\expandafter\let\csname endequation*\endcsname\relax 
\usepackage{amsmath}
\usepackage{color}

\usepackage{hyperref}
\hypersetup{
    colorlinks=true,
    linkcolor=blue,
    filecolor=magenta,
    urlcolor=blue,
}
\newcommand{\be}{\begin{equation}}
\newcommand{\ee}{\end{equation}}
\newcommand{\beal}{\begin{aligned}}
\newcommand{\eeal}{\end{aligned}}
\newcommand\bea {\begin{eqnarray}}
\newcommand\eea {\end{eqnarray}}
\newcommand{\bec}{\begin{cases}}
\newcommand{\eec}{\end{cases}}

\begin{document}

\title{Analogue black holes and scalar-dilaton theory}

\author[a]{Ian G. Moss}

\emailAdd{ian.moss@newcastle.ac.uk}

\affiliation[a]{School of Mathematics, Statistics and Physics, Newcastle University, 
Newcastle Upon Tyne, NE1 7RU, UK}

\date{\today}

\begin{abstract}{\noindent
This note analyses on the long wavelength dynamics of a two horizon analogue black hole system
in one spatial dimension. By introducing an effective  scalar-dilaton model we show that closed 
form expressions can be obtained for the time-dependent Hawking flux and the energy density of 
the Hawking radiation. We show that, in the absence superluminal modes, there is a vacuum instability. 
This instability is recognisable to relativists as the analogue to the destabilisation of the Cauchy 
horizon of a black hole due to vacuum polarization.
}
\end{abstract}

\maketitle

\section{Introduction}

Laboratory analogues of hawking radiation could provide insight into many of the
outstanding questions surrounding black hole evaporation 
\cite{Unruh:1980cg,Unruh:2007kqa,Barcelo:2005fc}. 
Bose-Einstein condensates (BEC) provide fertile ground for possible analogue models.
The long wavelength perturbations of a condensate are described by an effective scalar field
theory with a background metric. Given the appropriate background velocity field, the
analogue metric can have horizons that produce Hawking radiation in the form of
pseudo-quanta in the condensate field. This note focusses on one particular aspect of Hawking
radiation, which is the stability of the Hawking stress-energy at the Cauchy, or innermost horizon,
of a black hole.

Analogue models with two horizons are of particular interest because they were used
in the first experiments that produced Hawking radiation in the laboratory
\cite{Steinhauer_2014,Steinhauer:2015saa}. Early theoretical
investigations of the two-horizon system by Corley and Jacobson \cite{Corley:1998rk} 
revealed the possible existence of a laser
effect. If the effective scalar field theory has a dispersion relation of the form 
$\omega^2=k^2+k^4/\Lambda^2$, the Hawking radiation
outside the horizons grows exponentially, implying an instability of the vacuum state.
Later work, which did not rely on the WKB approximation used by Corley and Jacobson,
found a more nuanced situation with evidence for a laser effect depending on the particular set-up 
\cite{Jain_2007,Coutant:2009cu}.
Following Steinhauer's experiment, there have been a number of attempts to examine the 
laser phenomenon and reproduce the experimental results by solving the Gross-Piteavski 
equation for the mean condensate field with various forms of noise added to mimic the quantum 
fluctuations \cite{Tettamanti_2016,Steinhauer:2016hfa,Wang:2017wnj}. The results appear
to be rather dependent on the approach that is followed.

In this note, we focus only on the long wavelength dynamics of the two horizon system
in one spatial dimension. The methods used
are generalisations of the conformal methods introduced by Christensen and Fulling \cite{Christensen:1977jc}.
By introducing an effective  scalar-dilaton model we shall see that closed form expressions can be obtained
for the hawking flux and energy density of the Hawking radiation. 
We shall show that, even when we drop the superluminal modes, there is a vacuum instability. 
This instability is recognisable to relativists as the analogue to the destabilisation of the Cauchy 
horizon of a black hole due to vacuum polarization \cite{Birrell:1978th,Davies:1989ew}.

The Cauchy horizon has been of long-standing interest in relativity.
Traversing the Cauchy horizon would allow views of the singularity
inside the black hole and signal a breakdown of the cosmic censorship principle.
Instability of the Cauchy horizon would in this respect be a desirable thing
\cite{Poisson:1990eh,Brady:1998au},
but there has been a recent reawakening of interest in the issue and the possibility
that physical objects could traverse the Cauchy horizon unscathed \cite{Mallary:2018hfd,Burko:1995uq}.

The results below describe in detail how energy accumulates on the 
analogue Cauchy horizon. The analysis is limited to long wavelength modes, but
allows for density and sound speed variations. Whether these short wavelength modes regularise 
the energy density at the analogue Cauchy horizon is an interesting question for future work.
Hopefully, the analysis below may help guide future work on the mode analysis
that covers the dynamics on all scales.

\section{The spacetime geometry}

Analogue gravity is based on the fact that the velocity potential for waves in an Eulerian fluid
satisfy the relativistic wave equation with an analogue metric, the Unruh metric \cite{Unruh:1980cg}. 
In one dimension (1D),
\begin{equation}
ds^2={a\over c}\left\{
-(c^2-v_f^2)dt^2-2v_f dtdr+dr^2\right\},
\end{equation}
where $v_f(r)$ is the background fluid velocity and $c(r)$ is the sound speed. 
(It is convenient to take $v_f<0$ and $c>0$.)
The extra factor $a(r)$ is the density in the Unruh metric, but in 1D the wave equation is independent of $a$ and we
can choose $a$ as we wish.

The metric diagonalises using the time coordinate $\tau=t+r_*$, where
\begin{equation}
dr_*={v_f dr\over c^2-v_f^2},\label{rdownstar}
\end{equation}
Then
\begin{equation}
ds^2=a\left\{
-B(r)d\tau^2+B(r)^{-1}dr^2\right\}
\end{equation}
where
\begin{equation}
B(r)={c^2-v_f^2\over c}.
\end{equation}
Black hole horizons occur at the roots $r_i$ of $B(r)=0$.

\begin{figure} [htb]
\centering
\includegraphics[width=0.3\linewidth]{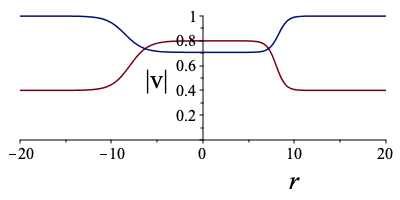}
\includegraphics[width=0.3\linewidth]{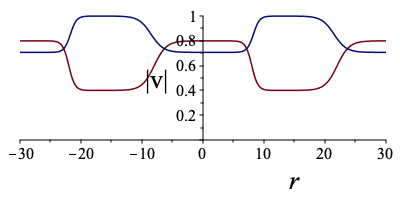}
\caption{A two-horizon fluid flow $v_f$ and the sound speed $c$ with two horizons (left) and a periodic
flow (right) that represents a circular topology.
} \label{fig:Flow}
\end{figure}

\begin{figure} [htb]
\centering
\includegraphics[width=0.25\linewidth]{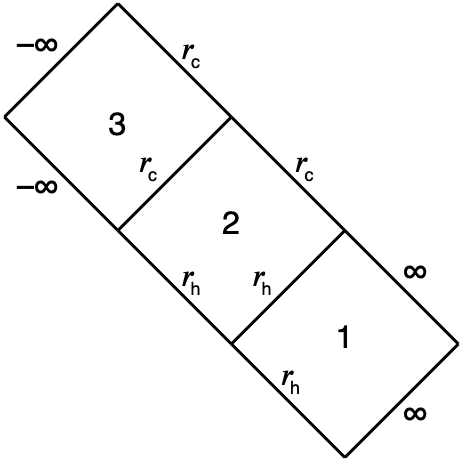}
\includegraphics[width=0.25\linewidth]{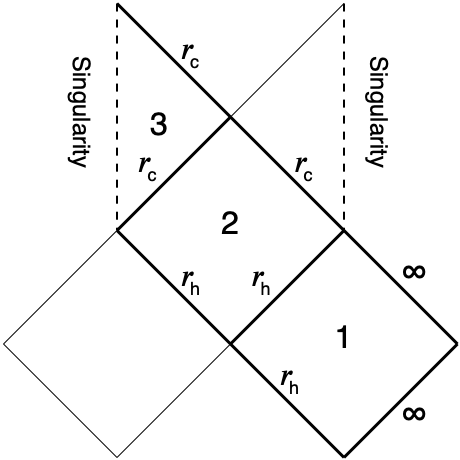}
\caption{Penrose diagrams for the analogue black hole (left) and the rotating black hole (right). 
The horizons are labelled $r_h$
and $r_c$. In region 2 of the analogue hole, $r_c<r<r_h$, right moving waves are left-moving in the 
original coordinates and so they end up dragged to the horizon $r_c$.
} \label{fig:FPenrose1}
\end{figure}

\begin{figure} [htb]
\centering
\includegraphics[width=0.18\linewidth]{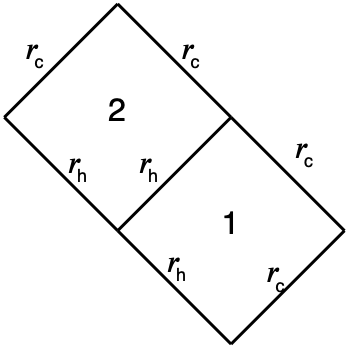}
\includegraphics[width=0.25\linewidth]{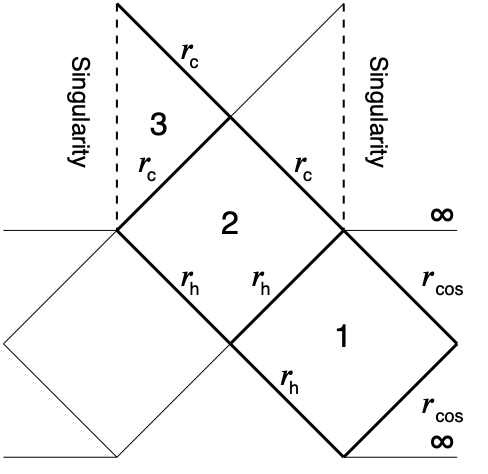}
\caption{Penrose diagrams for the analogue black hole with a circular topology (left) and the 
rotating black hole in de Sitter space (right). The latter has an additional horizon, the cosmological horizon,
at $r=r_{cos}$. In region 1 of the analogue hole, right moving waves travel round the circle and
meet the horizon at $r_c$. The topology of regions 1 and 2 are the same in both cases if we 
unwind the circle and work with the periodic potential.
} \label{fig:FPenrose2}
\end{figure}

The Penrose diagram for this spacetime is constructed from null Kruskal coordinates
as follows. 
First, define a new stretched coordinate $r^*$,
\begin{equation}
dr^*={c \,dr\over c^2-v_f^2}\label{rupstar}
\end{equation}
There are singularities in $r^*$ at the horizons, but this definition can be used for 
the entire range of $r$ providing we use a principle value prescription. Let $\kappa_i=B'(r_i)/2=c'+v_f'$,
then $r^*\to -\infty$ at any horizon where $\kappa_i>0$ and $r^*\to \infty$ at any
horizon where $\kappa_i<0$. In the region near the horizon, we define
\begin{align*}
U_i&=-e^{-\kappa_i(\tau-r^*)}\\
V_i&=e^{\kappa_i(\tau+r^*)}
\end{align*}
These null coordinates vanish at the horizon $r_i$ and the metric in regular there. They allow us
to extend the metric across the horizon. In the Penrose diagram,
lines of constant $U_i$ and $V_i$ have slope $\pm 45^\circ$ respectively.
The full Penrose diagram is constructed by patching these regions together to cover the entire range of $r$.

The figures show the Penrose diagram for a flow which is superluminal between
two fixed horizons $r=r_c$ and $r=r_h$, of opposite surface gravity. There are two cases,
a linear one with $-\infty<r<\infty$, and a circular topology shown as a periodic potential in the figure.
The horizon at $r_c$ is a Cauchy horizon for region 2, i.e. wave motion in region 3 cannot be predicted 
from initial data in region 2, although $r=r_c$ is not a Cauchy horizon for the full spacetime.
Note that this horizon is not a white hole, because null geodesics exit
a white hole and they fall into the horizon at $r=r_c$.

Figure \ref{fig:FPenrose1} shows the Penrose diagram of a four dimensional rotating black hole for
comparison. Each point in the Penrose diagram for the black hole represents a sphere, but
the topology of the space spanned by the (compactified) Kruskal coordinates for regions 1 and 2
is very similar. The differences lie mostly in the future extensions of the spacetime beyond
the Cauchy horizons. Is seems reasonable to regard the horizon at $r_c$ as an analogue
to the Cauchy horizon of a black hole.

\section{From BEC to scalar-dilaton theory}

This part of the story is not new and is only told is outline \cite{Michel:2016tog}. 
First, there is the action for the mean BEC field $\psi(r,t)$,
\begin{equation}
S=\int\left\{\frac{i}{2}\hbar\overline\psi\partial_t\psi-\frac{i}{2}\hbar\psi\partial_t\overline\psi
-{\hbar^2\over 2m}\partial_r\overline\psi\partial_r\psi-\frac12g(\overline\psi\psi)^2+
\mu\overline\psi\psi-V\overline\psi\psi \right\}drdt
\end{equation}
The trapping potential drives a background flow in the condensate. Potentials
in the transverse directions (not shown here) confine the condensate so that
it behaves as a 1D system.
The action is expanded about the time independent background, which has 
velocity $v$ and sound speed $c$,
\begin{equation}
\psi=\sqrt{\rho}e^{i\theta}(1+\delta\psi),\qquad v=\hbar\partial_r\theta/m,
\qquad g\rho=mc^2
\end{equation}
At quadratic order, let $\Psi=(\psi,\overline\psi)^T$, then the perturbed action reads
\begin{equation}
S[\Psi]=\int\rho \left\{ \frac{i}{2}\hbar\Psi^\dagger \sigma_z(\partial_t+v\partial_r)\Psi
+{\hbar^2\over 4m}\Psi^\dagger\partial_r^2\Psi-\frac12\Psi^\dagger{\cal H}\Psi \right\}drdt
\end{equation}
where $\sigma_i$ are the Pauli matrices and
\begin{equation}
{\cal H}=g\rho\sigma_x+g\rho 1.
\end{equation}
The field equation for the backgrounds $\rho$ and $v$ have been used. Next, diagonalise the
matrix ${\cal H}$, by introducing the normalised eigenvectors $u$ and $u'$, 
such that $\sigma_z u=u'$. Introduce normal modes $\varphi$ and $p$,
\begin{equation}
\Psi=-i\varphi u+\hbar^{-1} p u'
\end{equation}
In these fields,
\begin{equation}
S[\varphi,p]=\int \rho\left\{
p(\partial_t+v\partial_r)\varphi-\frac12{\hbar^2\over 2m}(\partial_r\varphi)^2
-\frac1{2\hbar^2}\left({\hbar^2\over 2m}(\partial_rp)^2+2mc^2p^2\right)
\right\}drdt
\end{equation}
Finally, consider the long wavelength limit where the field $p$ is composed mainly of
Fourier modes with $k\ll \xi^{-1}$, where the healing length $\xi=\hbar/mc$.
For consistency, we should see if this is consistent with the Hawking radiation. 
The Hawking spectrum is in the desired range if $k_BT_H\ll \hbar c/\xi$.
The formula for the Hawking temperature (see later) is $k_BT_H\sim \hbar v_f'$,
which implies that the background velocity field should not vary very much on
the scale of the healing length. This not unreasonable in actual experiments.

In the long wavelength limit, $p\approx \hbar^2(2mc^2)^{-1}(\partial_t+v\partial_r)\varphi$,
and eliminating $p$ gives
\begin{equation}
S[\varphi]={\hbar^2\over 2m}\int\rho\left\{
{1\over 2 c^2}[(\partial_t+v\partial_r)\varphi]^2-\frac12(\partial_r\varphi)^2
\right\}drdt
\end{equation}
Compare this to the fluid metric, 
\begin{equation}
g^{-1}|g|^{1/2}=c\left\{-{1\over c^2}({\bf e}_t+v{\bf e}_r)\otimes({\bf e}_t+v{\bf e}_r)
+{\bf e}_r\otimes{\bf e}_r\right\}
\end{equation}
The action therefore has a pseudo-geometrical form,
\begin{equation}
S[\varphi]={\hbar^2\over 2m}\int 
\left\{
-\frac12g^{\mu\nu}(\partial_\mu\varphi)(\partial_\nu\varphi)|g|^{1/2}
\right\}{\rho\over c}drdt
\end{equation}
In three spatial dimensions, the factors after the bracket are absorbed by the inverse metric.
However, in 1D this is not possible and the $\rho/c=(\rho m/g)^{1/2}$ term is always present. 
The system can be viewed instead as a model with an external dilaton field $\Phi$,
\begin{equation}
\Phi=-\frac14\ln\left({\hbar^2 \rho\over 4mg}\right).
\end{equation}
Then
\begin{equation}
S[\varphi]=-\hbar\int 
\left\{
\frac12g^{\mu\nu}(\partial_\mu\varphi)(\partial_\nu\varphi)e^{-2\Phi}
\right\}|g|^{1/2}drdt.
\end{equation}
This is still a conformally invariant model, i.e. the field equations are invariant under
rescaling of the metric and we can make an arbitrary choice of the factor 
$a$ in the metric. Different choices would, however, have an effect on the
energy density (see later).

In the $(\tau,r^*)$ coordinate system, the field equation for $\phi=e^{-\Phi}\varphi$ becomes
\begin{equation}
\partial_\tau^2\phi-\partial_{r^*}^2\phi+(\rho^{-1/4}\partial_{r^*}^2 \rho^{1/4})\phi=0
\end{equation}
Note that this equation is scattering problem with potential $\rho^{-1/4}\partial_{r^*}^2 \rho^{1/4}$.
The Hawking flux from a single horizon in this model has a grey-body spectrum with transmission
coefficients determined by the scattering potential.
However, the potential is very small in flows where $\rho$ only varies when close to the horizon.

\section{Scalar-dilaton theory}

 Much is known about scalar-dilaton theory in 2D because it was used to
 describe the back-reaction of Hawking radiation on the black hole spacetime
 \cite{Callan:1992rs,Russo:1992ht,Balbinot:1998yh,Kummer:1999zy} (the CGHS model). 
An effective action approach
similar to the one used below has been used for analogue black holes 
\cite{Balbinot:2004da,Balbinot:2004dc},
but previous work has only been applied to a single horizon and without the dilaton field.
 
First off, note that the stress-energy tensor of the scalar $\varphi$
is not conserved because of the external dilaton field $\Phi$. Fortunately,
the action is in geometric form and we can apply general covariance rules. Consider an
infinitesimal diffeomorphism $\delta g_{\mu\nu}=2\xi_{(\mu;\nu)}$ and
$\delta \Phi=\Phi^{;\mu}\xi_\mu$, then covariance implies
\begin{equation}
\delta S=\int\left\{2{\delta S\over\delta g_{\mu\nu}}\xi_{(\mu;\nu)}+
{\delta S\over\delta\Phi}\Phi^{;\mu}\xi_\mu\right\}d\mu=0
\end{equation}
Hence
\begin{equation}
\nabla_{\mu}T^\mu{}_\nu={\delta S\over \delta\Phi}\nabla_\nu\Phi
\end{equation}
In the quantum theory, this becomes an operator equation and we have
\begin{equation}
\nabla_{\mu}\langle T^\mu{}_\nu\rangle={\delta \Gamma\over \delta\Phi}\nabla_\nu\Phi
\end{equation}
where $\Gamma[g_{\mu\nu},\Phi]$ is the effective action with the external metric $g_{\mu\nu}$ and dilaton field $\Phi$.
 
The most important result for completing the theoretical description is the trace anomaly.
This was mired in controversy for a while, but in the end the correct result was given by Dowker \cite{Dowker:1998bm}
 \begin{equation}
 \langle T^\mu{}_\mu\rangle={\hbar\over 24\pi}\left(R-6\Phi_{;\mu}\Phi^{;\mu}+4\Phi_{;\mu}{}^\mu\right)
 \end{equation}
Bousso and Hawking demonstrated that it is possible to write down an effective action which is 
consistent with the trace anomaly \cite{Bousso:1997cg}. After correcting the trace anomaly,
 \begin{equation}
 \Gamma=-{\hbar\over 48\pi}\int\left\{\frac12R\square^{-1}R-6\Phi_{;\mu}\Phi^{;\mu}\square^{-1}R+4\Phi R\right\}d\mu
 \end{equation}
 The inverse d'Alembertian $\square^{-1}$ is less problematic than may first appear. For example,
 in the fluid metric $R=-\square\ln|aB|$, and we take the simple choice $\square^{-1}\square=1$.
 This effective action has been used extensively in discussions of the back reaction of Hawking
 radiation, but it should be noted that this action is not an exact result. Nevertheless, we continue with this action
 as in previous work. The functional derivative
  \begin{equation}
 {\delta\Gamma\over\delta\Phi}=-{1\over 12\pi}R-{1\over 4\pi}\left(\Phi^{;\mu}\square^{-1}R\right)_{;\mu}
 \end{equation}
We now have a complete set of equations that can be solved for the stress tensor $\langle T^\mu{}_\nu\rangle$.

Before moving on, consider energy conservation in this model. Suppose there is a
symmetry along the timelike vector $k^\mu$, i.e.
\begin{equation}
k_{\mu;\nu}+k_{\nu;\mu}=k^\mu\Phi_{;\mu}=0
\end{equation}
It follows that $\nabla_\mu(k^\nu T^\mu{}_\nu)=0$, and the corresponding conserved charge
is
\begin{equation}
H=-\int_\Sigma T^\mu{}_\nu k^\nu n_\mu dS,
\end{equation}
where $n_\mu$ is the normal to the surface $\Sigma$.
For the Unruh metric, $k^t=1$, $n_t=(ac)^{1/2}$ and $dS=(a/c)^{1/2}dr$,
\begin{equation}
H=-\int_\Sigma a\,T^t{}_t dr
\end{equation}
It is possible to verify that this integral is equal to the Hamiltonian of the model, and the energy density is $-a\,T^t{}_t$.

\section{Fluxes}

In this section we solve equations for the stress energy tensor of the Hawking
radiation in the long-wavelength limit. In two dimensions, the three components of stress energy, namely
the energy density, pressure and flux can be determined from the two conservation
laws and the trace anomaly, the latter being the only place where quantum field
theory is used \cite{Christensen:1977jc}.

It is convenient to work in the $(\tau,r)$ coordinate frame where the metic is diagonal.
We introduce the energy density $E(\tau,r)$, pressure $P(\tau,r)$ and flux $F(\tau,r)$,
\begin{equation}
E=-\langle T^\tau{}_\tau\rangle,\qquad P=\langle T^r{}_r\rangle,\qquad F=-\langle T^r{}_\tau\rangle.
\end{equation}
The energy and momentum conservation law
$\nabla_\mu \langle T^\mu{}_\nu\rangle=(\delta \Gamma/\delta\Phi)\nabla_\nu\Phi$ becomes
\begin{align}
a\dot E&=-(aF)'\\
aB^{-1}\dot F&=-(aBP)'+\frac12(aB)'(-E+P)+(aB)\Phi'{\delta\Gamma\over\delta\Phi}
\end{align}
Note that $F$ satisfies the usual flux conservation law and $E$ is the true energy 
density if we choose $a=1$. For the present, we remain agnostic on the choice of $a$.

The trace anomaly
$T^\mu{}_\mu=-E+P=q(R-6\Phi_{;\mu}\Phi^{;\mu}+4\Phi_{;\mu}{}^\mu)$, 
where $q=\hbar/24\pi$.
For our metric
\begin{equation}
R={Ba^{\prime 2}\over a^3}-{Ba''\over a^2}-{a'B'\over a^2}-{B''\over a},
\end{equation}
We use the trace anomaly to eliminate the energy density from the equations.
Remarkably, the right side of the momentum equation is an exact derivative, and the
equations become
\begin{align}
aB\dot P&=-B(aF)'\\
a\dot F&=-B(aBP-f)'
\end{align}
where
\begin{equation}
f(r)=-\frac{q}4 B^{\prime 2}-\frac{q}2 {BB'(a\rho)'\over a\rho}-\frac{q}4  {B^2 a^{\prime 2}\over a^2}
-\frac{q}2  {B^2 a'\rho'\over a\rho}
+\frac{3q}{16}{B^2\rho^{\prime 2}\ln|aB|\over \rho^2}
\end{equation}
Noting that $B\partial_r=\partial_{r^*}$, the equations combine into a wave equation
\begin{equation}
\partial_\tau^2(aBP-f)-\partial_{r^*}^2(aBP-f)=0
\end{equation}
Hence the general solution is
\begin{equation}
aBP=\Phi(v)+\Psi(u)+f(r)
\end{equation}
where $u=\tau-r^*$ and $v=\tau+r^*$. Substituting back for $F$,
\begin{equation}
aF=\Psi(u)-\Phi(v)
\end{equation}
and for the energy
\begin{equation}
aBE=\Phi(v)+\Psi(u)+h(r)
\end{equation}
where $h(r)=f(r)-aB\langle T^\mu{}_\mu\rangle$,
\begin{equation}
h(r)=-\frac{q}4 B^{\prime 2}-\frac{q}2 {BB'(a\rho)'\over a\rho}-\frac{5q}4  {B^2 (a\rho)^{\prime 2}\over (a\rho)^2}
+q{B^2(a\rho)''\over a\rho}+\frac{5q}8{B^2\rho^{\prime 2}\over\rho^2}+ qBB''
+\frac{3q}{16}{B^2\rho^{\prime 2}\ln|aB|\over \rho^2}.
\end{equation}

Finally, we want expressions for the energy and fluxes in the physical $t,r$ coordinate system.
Let $\beta=v_f/c$, then
\begin{align}
-T^t{}_t=&E-\beta B^{-1} F=-{\beta-1\over aB}\Psi+{\beta+1\over aB}\Phi+{h\over aB}\\
T^r{}_r=&P-\beta B^{-1} F=-{\beta-1\over aB}\Psi+{\beta+1\over aB}\Phi+{f\over aB}\\
-T^r{}_t=&F={1\over a}\Psi-{1\over a}\Phi
\end{align}
These are exact, closed expressions for the quantum stress tensor in the long wavelength
limit. However, in practice,
we have to integrate (\ref{rdownstar}) and (\ref{rupstar}) to obtain $u$ and $v$, 
so some numerical computation is necessary.

\subsection{Boundary conditions for the two-horizon case}

So far there are two functions $\Psi$ and $\Phi$ which depend on the initial conditions, but also
are constrained by regularity conditions at the horizons. In the two-horizon case with linear topology, 
there are three distinct regions with different coordinate charts for $u$. The solution for
$\Psi$ in region $i$ will be denoted by $\Psi_i(u)$. The $v$ coordinate has the same form in each of 
the regions, and there is a single function $\Phi(v)$.

At the horizons, $\beta\to -1$, $B\to 0$ and $h\to -q\kappa^2$. Regularity of the energy density $T^t{}_t$ requires that
\begin{align}
\hbox{at }r=r_h:&&\Psi_1(\infty)=\frac12q\kappa_h^2&&
\Psi_2(\infty)=\frac12q\kappa_h^2\\
\hbox{at }r=r_c:&&\Psi_2(-\infty)=\frac12q\kappa_c^2&&
\Psi_3(-\infty)=\frac12q\kappa_c^2
\end{align}
These relations are sufficient to show that the energy density will increase without limit near the
Cauchy horizon whatever the initial conditions.

\subsubsection{Equal surface gravities}

\begin{figure} [htb]
\centering
\includegraphics[width=0.3\linewidth]{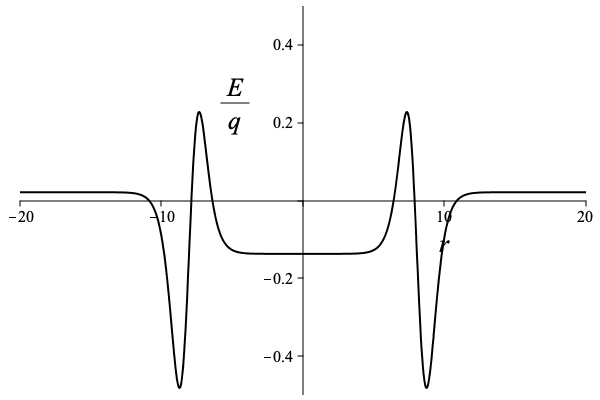}
\caption{Energy density of the Hawking radiation for the two horizon case, with $|\kappa_c|=|\kappa_h|$.}
\label{fig:StaticFlux}
\end{figure}

The conditions on $\Psi_2$ imply that a static solution with $\Psi_2$ constant can
only be achieved when the surface gravities have the same magnitude. In this case,
setting $\Phi=0$ and $a=1$ and $q=\hbar/24\pi$ gives the expected Hawking flux 
$\kappa^2/48\pi$ for a massless scalar field. We have
\begin{align}
-T^t{}_t&=-{\beta-1\over 48\pi B}\kappa^2+{h\over B}\\
T^r{}_r&=-{\beta-1\over 48\pi B}\kappa^2+{f\over B}\\
-T^r{}_t&={1\over 48\pi}\kappa^2
\end{align}
For a circular topology, the constant flux winds around the circle. For the linear topology,
we have the unrealistic situation with the flux extending to infinity.

A plot of the energy density shows considerable enhancement around each horizon
\ref{fig:StaticFlux}, compared to the asymptotoic value.

\subsubsection{Zero flux initial condition}

The initial conditions depend in detail on how the experiment is set up. We will
consider the case where there is no Hawking flux and no pressure at $t=0$.
The initial conditions are then
\begin{align}
\Psi_i(u)&=-\frac12 f(r)\hbox{ at $t=0$}\\
\Phi(v)&=-\frac12 f(r)\hbox{ at $t=0$}.
\end{align}
The simplest way of finding explicit expressions for the Hawking flux
is to convert the exact slutions back into
partial differential equations. Using  $u=t+r_*-r^*$,
\begin{align}
\partial_t\Phi&=(c-v_f)\partial_r\Phi,\\
\partial_t\Psi_i&=-(c+v_f)\partial_r\Psi_i.
\end{align}
After solving these equations, the functions $\Psi_i$ and $\Phi$ are substituted back into 
the exact solutions.

\begin{figure} [htb]
\centering
\includegraphics[width=0.3\linewidth]{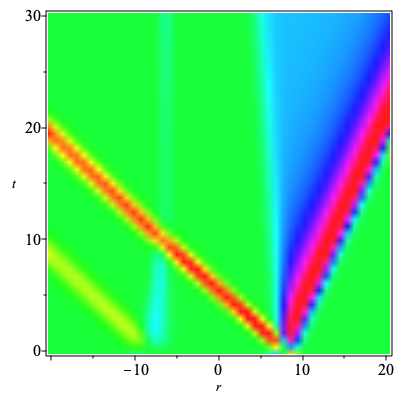}
\includegraphics[width=0.3\linewidth]{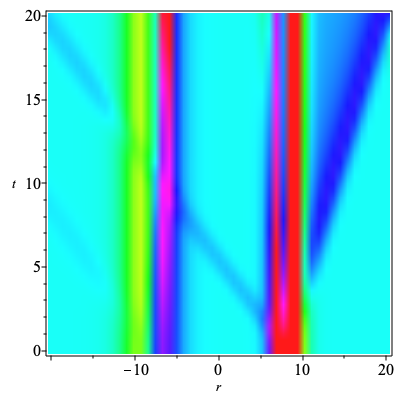}
\caption{Hawking radiation for the two horizon case, with $|\kappa_c|<|\kappa_h|$
and zero initial flux. The distance unit is the healing length of the sub-luminal region.
The Hawking flux is the blue patch emerging from the event horizon (left).
The energy density is plotted (right) and shows the concentration around the horizons, although the
instability shows up better in the next set of plots.
}
\label{fig:Flux}
\end{figure}

\begin{figure} [htb]
\centering
\includegraphics[width=0.3\linewidth]{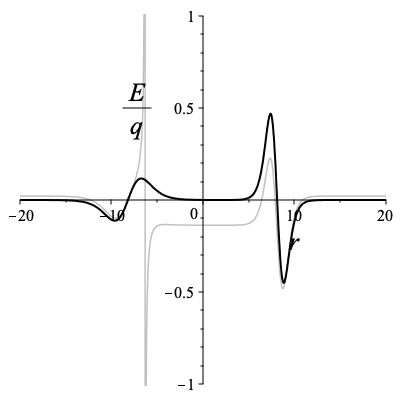}
\includegraphics[width=0.3\linewidth]{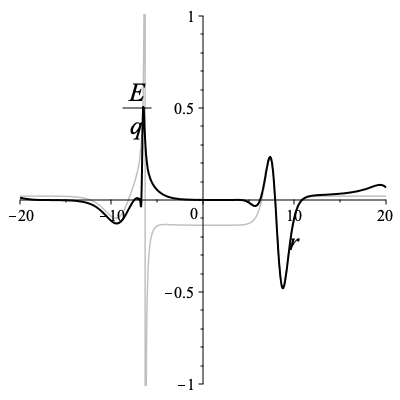}
\caption{
The Cauchy horizon instability is shown developing in the energy density from $t=0$ (left) to $t=20$ (right).
The energy density in the sub-luminal regions approaches the static result shown in grey.}
\label{fig:Flux}
\end{figure}

\section{Conclusion}

We have seen how the energy accumulates on the analogue Cauchy horizon
in the long wavelength approximation to waves on a BEC. As the energy builds up, 
the short wavelengths become more important. One interpretation of the black hole laser effect
is that superluminal modes carry away the growing energy from the Cauchy horizon.
Whether these short wavelength modes regularise 
the energy density at the analogue Cauchy horizon is an interesting question for future work.

\bibliography{Flux2D}

\end{document}